\documentclass[journal]{IEEEtran}
\usepackage{glossaries}
\usepackage{tabularx, booktabs}
\usepackage{multirow}
\usepackage{amssymb}
\usepackage{graphicx}
\usepackage{cases}
\usepackage{amssymb,amsmath,dsfont,mathrsfs}
\usepackage{color}
\usepackage{soul}
\usepackage{listings}
\usepackage[caption=false,font=footnotesize]{subfig}
\usepackage{setspace}
\usepackage{todonotes}
\usepackage{textcomp}
\usepackage{vwcol}
\usepackage{nicefrac}
\usepackage{phonetic}
\usepackage{pgf,tikz}
\usetikzlibrary{arrows}

\usetikzlibrary{pgfplots.groupplots}
\usetikzlibrary{positioning}
\usepackage{pgfplots}
\usetikzlibrary{plotmarks}
\usetikzlibrary{pgfplots.polar}
\usepgfplotslibrary{colorbrewer}

\usepackage[american]{babel}
\usepackage{multicol}
\usepackage[htt]{hyphenat}

\usepackage{algorithm}
\usepackage[noend]{algpseudocode}

\usepackage{xargs}
\usepackage{hyperref}
\usepackage{cleveref}
\crefname{equation}{}{}

\definecolor{mycolor1}{rgb}{0.01430,0.01430,0.01430}%
\definecolor{mycolor2}{rgb}{0.10740,0.22340,0.49840}%
\definecolor{mycolor3}{rgb}{      0,    0.3977,    0.4220}
\definecolor{mycolor4}{rgb}{0.00000,0.55650,0.24690}%
\definecolor{mycolor5}{rgb}{0.69120,0.77950,0.00000}%
\definecolor{mycolor6}{rgb}{0.96810, 0.89620,0.83780}
\definecolor{mycolor7}{rgb}{0.605469, 0.472656, 0.410156}%
\definecolor{mycolor8}{rgb}{0.605469, 0.472656, 0.410156}%

\definecolor{Mycolor1}{HTML}{396AB1}
\definecolor{Mycolor2}{HTML}{DA7C30}
\definecolor{Mycolor3}{HTML}{3E9651}
\definecolor{Mycolor4}{HTML}{CC2529}
\definecolor{Mycolor5}{HTML}{535154}
\definecolor{Mycolor6}{HTML}{6B4C9A}
\definecolor{Mycolor7}{HTML}{922428}
\definecolor{Mycolor8}{HTML}{948B3D}

\newacronym{cs}{CS}{compressed sensing}
\newacronym{pg}{PG}{proximal gradient}
\newacronym{ista}{ISTA}{iterative shrinkage-thresholding algorithm}
\newacronym{fbs}{FBS}{forward-backward splitting}
\newacronym{ama}{AMA}{alternating minimization method}
\newacronym{vmfb}{VMFB}{variable metric forward-backward}
\newacronym{fpg}{FPG}{fast proximal gradient}
\newacronym{zerofpr}{PANOC}{Proximal Averaged Newton-type algorithm for Optimality Conditions}
\newacronym{gn}{GN}{Gauss-Newton}
\newacronym{ip}{IPM}{interior point method}
\newacronym{scs}{SCS}{splitting conic solver}
\newacronym{ecos}{ECOS}{embedded conic solver}
\newacronym{mpc}{MPC}{model predictive control}
\newacronym{admm}{ADMM}{alternating direction method of multipliers}
\newacronym{drs}{DRS}{Douglas-Rachford splitting}
\newacronym{lasso}{LASSO}{least absolute shrinkage and selection operator}
\newacronym{fista}{FISTA}{fast iterative shrinkage-thresholding algorithm}
\newacronym{pc}{PC}{Pock-Chambolle algorithm}
\newacronym{fir}{FIR}{finite impulse response}
\newacronym{iir}{IIR}{infinite impulse response}
\newacronym{mp}{MP}{matching pursuit}
\newacronym{omp}{OMP}{orthogonal matching pursuit}
\newacronym{dft}{DFT}{discrete Fourier transform}
\newacronym{kcv}{KCV}{K-fold cross validation}
\newacronym{fbe}{FBE}{forward-backward envelope}
\newacronym{lbfgs}{L-BFGS}{limited memory BFGS}
\newacronym{sdp}{SDP}{semidefinite programs}
\newacronym{lti}{LTI}{linear time-invariant}
\newacronym{ir}{IR}{impulse response}
\newacronym{fft}{FFT}{fast fourier transform}
\newacronym{dct}{DCT}{discrete cosine transform}
\newacronym{gd}{GD}{gradient descent}
\newacronym{fao}{FAOs}{forward-adjoint oracles}
\newacronym{dag}{DAG}{directed acyclic graph}
\newacronym{socp}{SOCP}{second-order cone programming}
\newacronym{dcp}{DCP}{disciplined convex programming}
\newacronym{nmse}{NMSE}{normalized mean squared error}
\newacronym{mm}{MM}{majorize-minimization}
\newacronym{nn}{NN}{neural network}
\newacronym{dnn}{DNN}{deep neural network}
\newacronym{cnn}{CNN}{convolutional neural network}
\newacronym{pca}{PCA}{principal component analysis}
\newacronym{svd}{SVD}{singular-value decomposition}
\newacronym{ood}{OOD}{out-of-distribution}
\newacronym{ind}{IND}{in-distribution}
\newacronym{asr}{ASR}{automatic speech recognition}
\newacronym{fpr}{FPR}{false positive rate}
\newacronym{tpr}{TPR}{true positive rate}
\newacronym{de}{DE}{detection error}
\newacronym{auroc}{AUROC}{Area Under the Receiver Operating Characteristic curve}
\newacronym{aupr}{AUPR}{Area Under the Precision-Recall curve}
\newacronym{odin}{ODIN}{out-of-distribution detector for neural networks}
\newacronym{kmnist}{KMNIST}{Kuzushiji-MNIST}
\newacronym{fmnist}{FMNIST}{Fashion-MNIST}

\renewcommand{\vec}[1]{\mathbf{#1}}

\newcommand*\Real{\mathds{R}}
\newcommand*\tr{^{\intercal}}
\newcommand*\mean{\boldsymbol{\mu}}
\newcommand*\cov{\boldsymbol{\Sigma}}
\newcommand*\icov{\cov^{-1}}

\newcommand\ie{%
	\@ifnextchar,{\textit{i.e.}}{\textit{i.e.}, }%
}
\newcommand\eg{%
	\@ifnextchar,{\textit{e.g.}}{\textit{e.g.}, }%
}

\begin{document}

\title{A $t$-distribution based operator for enhancing \\ out of distribution robustness of neural network classifiers}

\author{Niccol\`o Antonello, Philip N. Garner
  \thanks{
  The authors are with Idiap Research Institute, 
  Martigny, Switzerland.
  E-mail: nantonel@idiap.ch, pgarner@idiap.ch
  This research was supported by 
  Innosuisse 
  under the project 27674.1.PFES-ES-SHAPED.
  
  \copyright~2020 IEEE. 
  Personal use of this material is permitted.  
  Permission from IEEE must be obtained for all other uses, 
  in any current or future media, 
  including reprinting/republishing 
  this material for advertising or promotional purposes, 
  creating new collective works, 
  for resale or redistribution to servers or lists, 
  or reuse of any copyrighted component 
  of this work in other works.
}
}

\maketitle

\begin{abstract}
Neural Network (NN) classifiers can assign extreme probabilities
to samples that have not appeared during training (out-of-distribution samples)
resulting in erroneous and unreliable predictions.
One of the causes for this unwanted behaviour
lies in the use of the standard softmax operator which
pushes the posterior probabilities to be either zero or unity
hence failing to model uncertainty.
The statistical derivation of the softmax operator
relies on the assumption that
the distributions of the latent variables for a given class are Gaussian with known variance.
However, it is possible to use different assumptions in the same derivation
and attain from other families of distributions as well.
This allows derivation of novel operators
with more favourable properties.
Here, a novel operator is proposed that is
derived using $t$-distributions which are capable of
providing a better description of uncertainty.
It is shown that classifiers that adopt this
novel operator can be more robust to out of distribution samples,
often outperforming NNs that use the standard softmax operator.
These enhancements can be reached with minimal
changes to the NN architecture.
\end{abstract}

%\begin{IEEEkeywords}
%  TODO
%\end{IEEEkeywords}

\IEEEpeerreviewmaketitle

\section{Introduction}\label{sec:intro}

Numerous scientific fields have found 
in deep learning efficient tools for classification. 
When enough labelled data is available a 
\gls{nn} is nowadays the most popular 
choice and often represents the state-of-the-art.
%%%
Many signal processing tasks have recently  
adopted \gls{nn} classifiers as well, 
for example in computer vision~\cite{han2018advanced},  
audio processing~\cite{mesaros2017detection,purwins2019deep,bianco2019machine}
and speech recognition~\cite{hinton2012deep,xiong2017toward}.
%%%
Despite their popularity, \gls{nn} classifiers 
can sometimes completely fail in their predictions,
assigning high confidence to \gls{ood} samples~\cite{nguyen2015deep}.
This problematic behaviour has been 
addressed in a number of recent studies 
where different solutions have been proposed. 

A direct solution to this issue is train classifiers 
augmenting the training set with \gls{ood} samples. 
It has been shown that this exposure 
generalizes well to 
unmodeled distributions~\cite{hendrycks2019deep,lee2018training}.
These \gls{ood} datasets can also be obtained 
by synthesising adversarial samples~\cite{lee2018training,hein2019relu}. 
However, there can never be the certainty that these \gls{ood} 
datasets are representative of all the possible   
\gls{ood} samples that might be present in real case scenarios.
Moreover, these procedures can substantially increase 
the training time and often 
require additional tuning of hyperparameters.
Obtaining reliable classifiers 
without the need of employing \gls{ood} datasets 
is therefore attractive. 
This can be achieved
by developing robust {\em confidence measures} 
that can indicate whether the classifier is wrong. 
The simplest confidence measure that can be obtained 
without any modification of the \gls{nn} architecture
consists of the softmax output~\cite{Hendrycks2016}, 
although we argue that this can often be unsatisfactory.
More trustworthy confidence measures can be produced through 
Monte Carlo methods for example   
by employing dropout during the forward pass~\cite{Gal2016}. 
Alternatively, confidence measures can be obtained 
by combining the outputs of an ensemble of classifiers~\cite{Vyas2018}. 
However, these approaches are rather computationally demanding.
More recently \gls{ood} identification has 
been achieved by a temperature scaling of 
the softmax operator combined with 
a perturbation of the input sample~\cite{liang2017enhancing}. 
This perturbation is obtained through differentiation
and can also be considered costly for certain applications.
Other techniques rely on learning confidence measures 
using particular cost functions and 
appending to \glspl{nn} architecture 
an additional output branch
that provides a score of reliability~\cite{Kendall2017,DeVries2018}. 

In this paper a novel method that 
produces reliable 
confidence measures is proposed. 
This does not require the need of any 
\gls{ood} distribution dataset, 
avoids any substantial increase of the computational complexity
and works with standard training procedures.
The key idea is to replace the softmax operator 
of \glspl{nn} with a different operator that 
inherently promotes low confidence for \gls{ood} samples.
This operator is obtained 
by a modification of the statistical derivation of the softmax operator
that amounts to using $t$-distributions instead of Gaussians
to model the probability of latent variables for a given class.
For this reason the proposed operator is 
named {\em $t$-softmax}. 
This idea can be traced back at least to~\cite{bedworth1992importance}
where it is shown that posterior probabilities 
obtained using $t$-distributions
instead of Gaussian distributions 
can avoid high confidence predictions  
in regions where data is not available. 
Variations of the softmax operator 
have already appeared
but for other purposes such as 
enhanced embeddings clustering~\cite{liu2017sphereface,wang2018additive}
and for reinforcement learning~\cite{asadi2017alternative}.

\begin{figure*}[t!]
  \centering
  \includegraphics{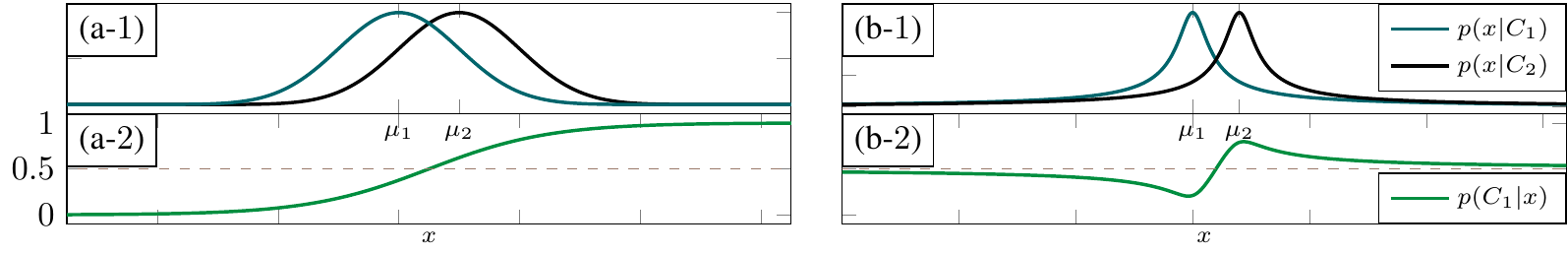}
  \caption{
    Sigmoid function (a-2) and $t$-sigmoid function (b-2) can be 
    derived as posterior probabilities $p(C_i | x)$.
    The conditional distributions used to model 
    the distribution for the latent variable $x$ 
    are either Gaussian distributions (a-1) 
    or $t$-distributions with $\nu = 1$ (b-1).
  }\label{fig:sigmoids}
\end{figure*}

\section{Statistical derivation of the softmax operator}
\label{sec:softmax}

The statistical derivation 
of the softmax operator can be found in~\cite{bridle1990probabilistic} 
and is summarized in this section.
Bayes's theorem can be used to express the probability 
of a sample $\vec{x} \in \Real^N$ from a random latent variable 
belonging to the class $i$ as:
\begin{equation}
  P(C_i | \vec{x} ) = 
  \frac{
    p(\vec{x} | C_i) P(C_i) 
    }{
    \sum_{i = 1}^{N_c} p(\vec{x} | C_i) P(C_i)
  }, 
  \label{eq:bayes}
\end{equation}
where $N_c$ is the number of classes and  
$C_i$ and $\vec{x}$ refer to the events that $C = i$ and
the sample takes value $\vec{x}$, respectively.
It is assumed that $p(\vec{x} | C_i)$  
is a Gaussian distribution 
with mean $\mean_i \in \Real^N$
and covariance matrix $\cov_i \in \Real^{N \times N} $: 
\begin{equation}
  p(\vec{x} | C_i) = 
  \tfrac{1}{\sqrt{(2 \pi)^N | \cov_i | }}
  \exp \left( 
    -\tfrac{1}{2} 
    (\vec{x}+\mean_i)\tr \icov_i (\vec{x}+\mean_i)
  \right),
  \label{eq:gauss}
\end{equation}
for $i = 1, \dots, N_c$.
Furthermore, it is assumed 
that the prior probabilities are all equal 
\ie~$P(C_1) = P(C_2) = \dots = P(C_{N_c})$
and that all Gaussian distributions share the same covariance matrix 
\ie $\cov_1 = \dots = \cov_{N_c}= \cov$. 
Under these assumptions, 
substituting~\eqref{eq:gauss} into~\eqref{eq:bayes} 
the {\em softmax operator} can be obtained:
\begin{equation}
  P(C_i | \vec{x} ) = 
  \frac{
    \exp \left( - \tfrac{1}{2} \left( 
    a +   \vec{w}_i\tr \vec{x} + b_i 
    \right) \right)
  }{
  \sum_{k=1}^{N_c} 
  \exp \left( - \tfrac{1}{2} \left( 
  a +  \vec{w}_k\tr \vec{x} + b_k 
  \right) \right)
}
\label{eq:softmax}
\end{equation}
where:
\begin{equation}
  a = \vec{x}\tr \icov \vec{x}, 
  \label{eq:quad_const}
\end{equation}
is a constant that vanishes and
\begin{equation}
  \vec{w}_i = 2 \icov \mean_i,
  \label{eq:weights}
\end{equation}
can be interpreted as the weights and 
\begin{equation}
  b_i = \mean_i\tr \icov \mean_i,
  \label{eq:bias}
\end{equation}
as the biases of a fully connected layer
$\vec{W} \vec{x} + \vec{b}$ with 
$\vec{W} = [\vec{w}_1; \dots; \vec{w}_{N_c}] \in \Real^{N_c \times N}$
and $\vec{b} = [ b_1  \dots  b_{N_c}  ]\tr$.
Modern \glspl{nn} utilize 
fully connected layers that do not require
the quadratic constraints~\cref{eq:weights,eq:bias} 
to be satisfied implying that 
this statistical interpretation 
is usually not valid. 
During training, $\vec{W}$ and $\vec{b}$ implicitly learn $\cov$.  
therefore here it can be assumed that $\cov = \vec{I}$
where $\vec{I}$ is the identity matrix.

\begin{figure}[t]
  \centering
  \includegraphics{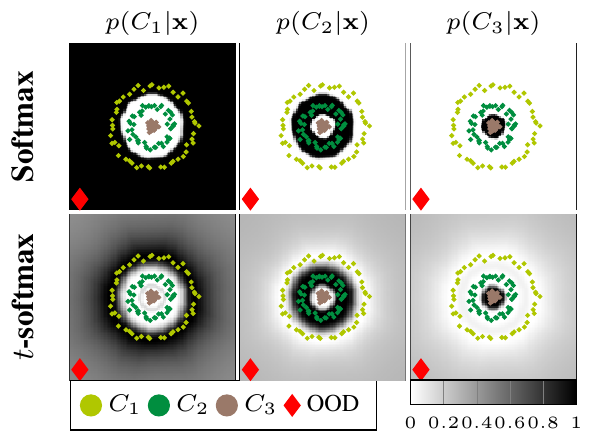}
  \caption{
    Toy example of a 3 class \gls{nn} classifier.
    Color visualizes the output probability $p(C_i | \vec{x})$
    on a fine grid.
    Top figures and bottom figures show the output of classifiers
    that use either softmax or $t$-softmax respectively.
  }\label{fig:simple_class}
\end{figure}

\Cref{fig:sigmoids}(a) visualizes
this derivation for the one-dimensional binary case.
Here, the softmax operator reduces to the so called 
{\em sigmoid function}. 
The graph shows how the sigmoid function 
tends to make mostly high confidence predictions
\ie~with $p(C_i | x)$ approaching either 0 or 1
for the largest part of its domain.
This behaviour is certainly positive 
for \gls{ind} samples, 
however it persists also when $x$  
is far from the means $\mu_1$ and $\mu_2$. 
Consequently, the sigmoid function 
may lead to erroneous predictions 
assigning high scores to \gls{ood} samples. 
This characteristic is also present
in the multi-dimensional case and complex 
\gls{nn} classifiers that employ 
the softmax operator as their output. 
The top scatter plots of \Cref{fig:simple_class} 
show a toy example of a classifier 
that separates two-dimensional features 
into three classes.
Here, the \gls{nn} classifier consists of three 
fully connected layers with ReLU nonlinearities. 
This is trained  
using stochastic gradient descent using simulated data.
As the top plots of \Cref{fig:simple_class} show, 
the three classes are correctly identified.
However when an \gls{ood} sample is fed to the \gls{nn}, 
totally erroneous predictions can be given.
For example in \Cref{fig:simple_class},
on the bottom left corner 
an \gls{ood} sample is depicted 
using a diamond marker. 
The classifier wrongly assigns this sample  
to the first class 
with a probability that equals to 1. 

\section{The $t$-softmax operator}
\label{sec:tsoftmax}

The bottom plots of \Cref{fig:simple_class}
show a more trustworthy classifier that does not 
assign high probability to \gls{ood} samples.
Here, the diamond marker \gls{ood} sample
is given a probability of 0.5 for the first class and 
0.25 for the second and third classes.
This means the uncertainty regarding this \gls{ood} 
sample can be directly assessed from the outputs
of the classifier which can then be considered 
reliable confidence measures.
Such a classifier is obtained by replacing 
the softmax operator layer with a different operator 
that can better model uncertainty.
The key idea is to employ {\em $t$-distributions} 
instead of Gaussian distributions for $p(\vec{x} | C_i)$: 
\begin{equation}
  p(\vec{x} | C_i) = 
  \tfrac{1}{\sqrt{\nu} B \left( \tfrac{1}{2}, \tfrac{\nu}{2} \right)}
  \left( 
    1+
    \tfrac{1}{\nu} 
    (\vec{x}+\mean_i)\tr (\vec{x}+\mean_i)
  \right)
  ^{ - \tfrac{\nu+1}{2} },
  \label{eq:tdistro}
\end{equation}
where $B$ is the Beta function and 
$\nu$ is a positive real number. 
The choice of $t$-distribution is not arbitrary; it arises as a Gaussian distribution~\cite[Sec.3.2]{gelman2013bayesian} where the variance is assumed unknown but distributed as inverse $\chi^2$ (parametrized by $\nu$).
The  $t$-distribution can be seen as a generalization of the Gaussian distributions, 
with~\eqref{eq:tdistro}
approaching~\eqref{eq:gauss} as $\nu$ increases. 
Indeed, the parameter $\nu$ controls
how fast $t$-distribution's tails decay.
While the tails of Gaussian distribution decay exponentially
those of $t$-distribution's ones decay polynomially, hence more slowly.
Such difference can be seen on the top plots of \Cref{fig:sigmoids}.
As a result, the $t$-distribution models 
uncertainty more effectively than Gaussian distributions 
and this can also be seen in the resulting operator 
named {\em $t$-softmax}, obtained    
substituting~\eqref{eq:tdistro} into~\eqref{eq:bayes}:
\begin{equation}
  P(C_i | \vec{x} ) = 
  \frac{
    \left( 
      1 + \tfrac{1}{\nu}
    \left( 
    a +   \vec{w}_i\tr \vec{x} + b_i 
    \right)
  \right)
    ^{ - \tfrac{\nu+1}{2} }
  }{
  \sum_{k=1}^{N_c} 
    \left( 
      1 + \tfrac{1}{\nu} 
    \left( 
    a +   \vec{w}_k\tr \vec{x} + b_k 
    \right)
  \right)
    ^{ - \tfrac{\nu+1}{2} }
}.
  \label{eq:tsoftmax}
\end{equation}
In contrast to the softmax operator, 
in~\eqref{eq:tsoftmax}
$a$ does not vanish and 
the quadratic constraints~\cref{eq:quad_const,eq:weights,eq:bias}
are required in order 
to avoid complex or negative outputs.
%%%
Unlike in the Gaussian case where choosing $\cov = \vec{I}$ 
in~\cref{eq:quad_const,eq:weights,eq:bias}
preserves rotational invariance, here this is no longer the case.
However, as the results will show this does not appear to
decrease the \gls{nn} classifiers accuracy.
%%%
It is possible to impose~\cref{eq:quad_const,eq:weights,eq:bias} by construction.
If $\vec{X} \in \Real^{N \times N_b}$
is an input consisting of $N_b$ batches then 
the dot products inside the parenthesis of~\eqref{eq:tdistro}
and expanded as $a +  \vec{w}_i\tr \vec{x} + b_i $
in~\eqref{eq:tsoftmax}
can be collected as: 
\begin{equation}
  \vec{Y} = \left( \vec{A} + \vec{W} \vec{X} + \vec{B} \right) \in \Real^{N_c \times N_b}. 
  \label{eq:quadratic_layer}
\end{equation}
The $(i,j)$-th elements of $\vec{A}$ and $\vec{B}$ 
are given by 
$a_{i,j} = \vec{x}_j\tr\vec{x}_j$ 
$b_{i,j}= 1/4 {\vec{w}}_i\tr{\vec{w}}_i = \mean_i\tr \mean_i$ 
for $i=1,\dots,N_c$ and $j=1,\dots,N_b$ 
with $\vec{x}_j$ being the $j$-th column of $\vec{X}$ and 
${\vec{w}}_i$ being the $i$-th row of $\vec{W}$.
Since all of the rows and columns 
of $\vec{A}$ and $\vec{B}$
are identical, respectively,
only $N_c+N_b$ dot products are needed 
to construct these matrices 
whose full storage is not needed.
Therefore, the layer defined in \eqref{eq:quadratic_layer}, 
named here as {\em quadratic layer},
does not require a substantial increase 
of computational resources when compared to 
a fully connected layer. 

\Cref{fig:sigmoids}(b-2) shows the 
resulting function for the two-dimensional binary case.
Unlike the sigmoid, this function
approaches 0.5 when far from $\mu_1$ and $\mu_2$
therefore promoting uncertainty 
on the \glspl{ood}.
Looking back at \Cref{fig:simple_class} 
it is clear that such behaviour 
is reflected in the \gls{nn} classifier 
that employs the $t$-softmax operator.
Wide areas surrounding  
the \glspl{ind} 
consistently give low confidence predictions.
Remarkably, even though the 
nonlinearity of this operator is stronger 
with respect to the softmax operator,
standard stochastic optimization methods 
can be employed using well known cost functions.

\section{Evaluation}\label{sec:results}

%%%%
In this section a number of experiments is presented 
to support the hypothesis that $t$-softmax can lead to 
classifiers that are more robust to \gls{ood} samples.
Additionally, it is hypothesized that $\nu=1$ 
should increase \gls{ood} robustness in most of the scenarios.
In a more thorough formulation, $\nu$ would be related 
to the dimension and to the prior on the variance of 
the latent variables; this is a matter for future research.
%%%%

\subsection{Baseline and state-of-the-art}\label{sec:baseline_sota}

The proposed method is compared with a baseline and 
a state-of-the-art method. 
The baseline was proposed in~\cite{Hendrycks2016} 
and simply consists
in using the maximum of the softmax output 
as a confidence measure.
%%%
Here, only standard fully connected layers are employed.
This means that the equivalence 
between the softmax and the $t$-softmax operators 
for large $\nu$ is not reached 
since~\cref{eq:weights,eq:bias} are not satisfied.
%%%
A more recent state-of-the-art method 
has been proposed in~\cite{liang2017enhancing}
and is known as \gls{odin}.
This method consists of adding the following perturbation 
to the input sample
$
\tilde{\vec{x}} = 
\vec{x} - \epsilon \text{sign} \left( 
-\nabla_{\vec{x}} \log \sigma (\vec{x}/\gamma) 
\right) 
$
where $\sigma$ is the softmax operator, 
$\epsilon$ is a small constant 
and $\gamma$ is a scaling factor.
%In our experiments $\epsilon$ and $\gamma$ 
%are tuned using Gaussian noise as a \gls{ood} distribution.
A confidence measure is then obtained by feeding 
$\tilde{\vec{x}}$ to the classifier 
and taking the maximum score of the scaled 
softmax output.
However, this technique is 
computationally demanding since it requires 
the gradient. 
Specifically, a forward-backward pass
and an additional forward pass are needed. 
This means \gls{odin} is at least 
three times more expensive than 
the baseline and the proposed method. 
%%%%
We verified this using the 
\gls{nn} described in~\Cref{sec:cnn}.
The average time per sample
using $t$-softmax operator and \gls{odin} 
is 0.44~ms and 2.58~ms respectively,
meaning that with the current implementation \gls{odin} 
is approximatively six times slower than the proposed method. 
%%%%

\subsection{Evaluation metrics}\label{sec:fom}

Following~\cite{liang2017enhancing}
these figures of merit are used for evaluation. 
\glsunset{fpr}\glsunset{tpr}
{\gls{fpr} at 95\% \gls{tpr}}
\glsreset{fpr}\glsreset{tpr}
gives a measure of the possibility of 
\gls{ood} sample being interpreted as 
an \gls{ind}, \ie~the \gls{fpr},
when the \gls{tpr} is 95\%.
\glsunset{de}
{Detection error (DE)} is the misclassification probability
when \gls{tpr} is 95\% and is given by 
$\text{DE} = 0.5 (1-\text{TPR}) + 0.5 \text{FPR}$.
It assumes \gls{ind} and \gls{ood} samples are 
equal in number. 
{AUROC} is the \gls{auroc}. The ROC curve~\cite{fawcett2006introduction} 
is the plot between \gls{tpr} and \gls{fpr} at different 
operating points of the classifier. Larger area under this curve 
implies more robust classifiers. 
{AUPR} is the \gls{aupr}. 
Similarly, the PR curve~\cite{saito2015precision}  
is the plot between precision ($\text{TP}/ (\text{TP+FP})$) and 
recall ($\text{TP}/(\text{TP}+\text{FN})$) 
at different operating points. 

\subsection{Result on FMNIST and KMNIST}\label{sec:cnn}

The proposed method is evaluated on two image classification 
tasks using \gls{fmnist}~\cite{fmnist} and 
\gls{kmnist}~\cite{kmnist} datasets. 
These are datasets of images of clothes and 
cursive Japanese Hiragana characters respectively 
and are both intended as a more challenging replacement 
of the well-known MNIST dataset~\cite{mnist}.
%In fact these datasets have identical specifics to MNIST, 
%having $7 \cdot 10^4$ gray scale $28 \times 28$ pixel images
%divided in 10 classes.
For these experiments a standard 
\gls{cnn}~\cite{lecun1989backpropagation} is used. 
Its architecture consists of two 
two-dimensional convolutional layers with  
20 and 50 channels respectively 
having two-dimensional max pooling 
and ReLU nonlinearities.
The convolutional layers are followed by 
three fully connected layers that 
successively reduce the latent variable dimensions from 
$800$ to $500$, $100$ and $10$.
ReLU nonlinearities are also 
employed between these layers.
When $t$-softmax is used the last fully connected layer
is replaced by the quadratic layer described in 
\Cref{sec:tsoftmax}.
%%%
Classifiers are trained using PyTorch~\cite{pytorch} 
in a reproducible manner~\cite{tsoftmaxcode}.
%%%
Specifically, stochastic gradient descent is used
with Nesterov acceleration, weight decay of $5 \cdot 10^{-4}$,
momentum of $0.5$, batches of 128 samples,
a learning rate of $0.1$ for $20$ epochs 
using cross entropy as loss function.

\begin{figure}[t]
  \centering
  \includegraphics{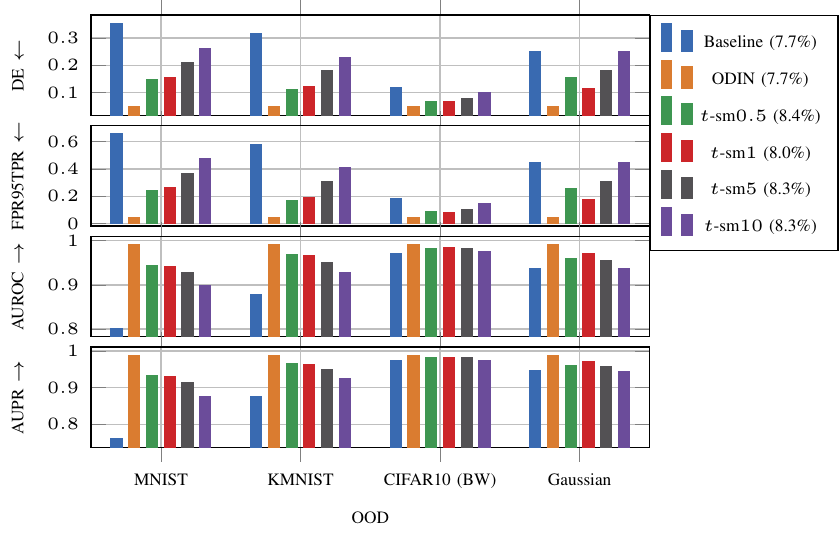}
  \caption{
    Results for a \gls{fmnist} classifier. 
    Arrows on the y-axis labels indicate if the 
    figure of merit should be either positive or negative 
    for the classifier to be more robust.
    In the legend, $t$-sm stands for $t$-softmax
    and the number next to the abbreviation is value of $\nu$.
    Percentages between parenthesis 
    in the legend indicate the test error.
  }\label{fig:fmnist}
\end{figure}

\begin{figure}[t]
  \centering
  \includegraphics{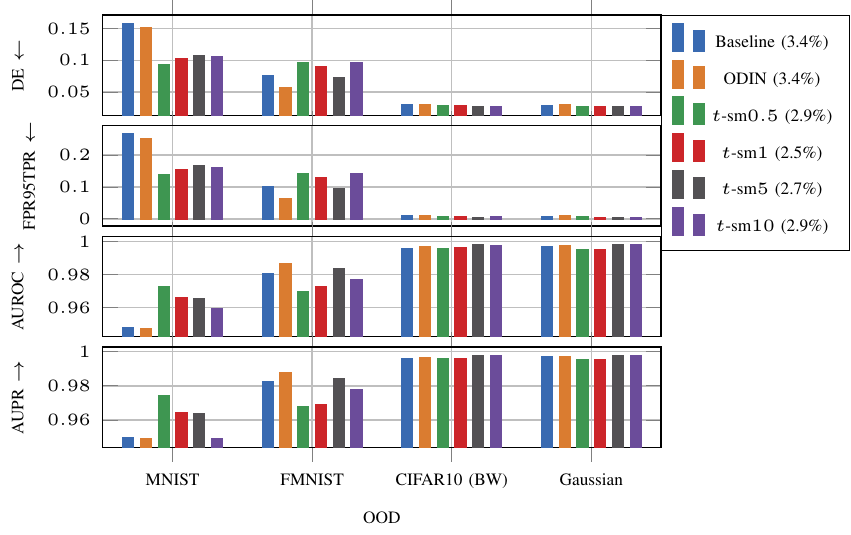}
  \caption{
    Results for a \gls{kmnist} classifier. 
  }\label{fig:kmnist}
\end{figure}

\Cref{fig:fmnist} shows plots of 
the different figures of merit 
for the \gls{fmnist} classifier. 
The classifiers that utilize $t$-softmax 
have been trained using different values of $\nu$ 
to check the effect of this parameter.
Looking at the percentages in the legend 
that indicate the test error of the classifiers,
similar performances are reached 
with slightly worse test errors 
for the classifiers with $t$-softmax.
The confidence measures are obtained  
from the \gls{ind} test dataset and 
different \gls{ood} datasets 
such as MNIST, \gls{kmnist}, gray scale CIFAR10~\cite{krizhevsky2009learning} 
and Gaussian noise.
An equal number of \gls{ind} 
and \gls{ood} samples is used.
These confidence measures are then used to 
evaluate the robustness of the classifiers using 
the figures of merit described in \Cref{sec:fom}.
The performance of \gls{odin} is superior, 
however $t$-softmax sometimes evens \gls{odin} 
for example when CIFAR10 is used as \gls{ood}. 
Similar results can be seen in \Cref{fig:kmnist}
that shows the same figures of merit 
for the \gls{kmnist} classifiers. 
Here \gls{fmnist} is also used as an \gls{ood} dataset.
It can be seen that in this case slightly better test error 
are achieved by the $t$-softmax classifiers.
However, the baseline is not surpassed 
by all $t$-softmax classifier configurations. 
Specifically, when \gls{fmnist} is the \gls{ood} dataset 
$t$-softmax has inferior figures of merit 
with respect to the baseline,
but not in the case $\nu=5$ where 
it reaches slightly better results.
Overall, \gls{odin} reaches results that are very close 
to the ones of the $t$-softmax classifiers.
\gls{odin} is superior only when \gls{fmnist} is the \gls{ood} dataset
but is inferior with MNIST.
%%%
In conclusion, it can be seen that all $t$-softmax classifiers 
consistently outperform the baseline,
with $\nu=1$ giving the overall best performances, 
confirming the hypothesis that this setting 
can be used in practice.
%%%

\subsection{Result on CIFAR10}\label{sec:cifar10}

\begin{figure}[t]
  \centering
  \includegraphics{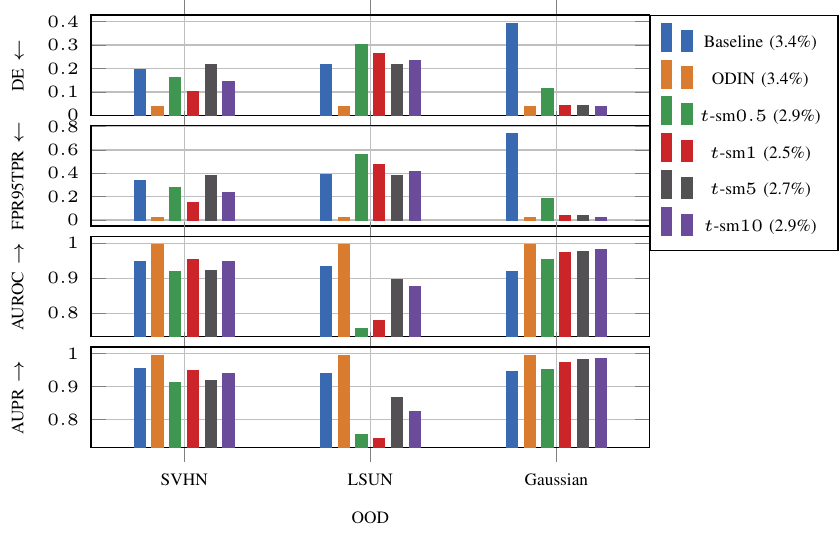}
  \caption{
    Results for a CIFAR10 classifier using Densenet. 
  }\label{fig:cifar10}
\end{figure}

A more advanced state-of-the-art \gls{nn}
known as Densenet~\cite{huang2017densely} with a depth of 100 
is trained on CIFAR10~\cite{krizhevsky2009learning}, a dataset 
consisting of coloured $32 \times 32$ pixel images 
of objects belonging to 10 different classes.
A similar training procedure as the one described in
\Cref{sec:cnn} is used,
with 300 epochs, batch size of 64 and a weight decay of $10^{-5}$.
Learning rate is decreased by a factor of 10 at the 
$150$th and $225$th epoch.
Here different \gls{ood} datasets are used: 
SVHN~\cite{netzer2011reading} and 
resized images of the LSUN test set~\cite{yu2015lsun}.
\Cref{fig:cifar10} shows the evaluation results.
The results are less convincing when 
compared to the previous experiments.
In particular, when the \gls{ood} dataset is the 
LSUN dataset $t$-softmax reaches performance that 
are below the baseline. 
%%%%
However, the same conclusions 
of \Cref{sec:cnn} are 
supported by the Gaussian \gls{ood} experiment 
where significant improvements are seen.
%%%%

\section{Conclusion}

In this paper a novel operator named $t$-softmax
is proposed.
It is shown that its statistical derivation  
is analogous to the one used for the softmax operator
with the only difference that $t$-distribution are 
used to better model uncertainty.
Results on image classification tasks 
show that classifiers trained using $t$-softmax 
can be robust to \gls{ood} samples. 
The proposed method is sometimes comparable to  
a state-of-the-art method named \gls{odin}
which is at least
three times more computationally expensive.
%Preliminary results using state-of-the-art 
%\gls{nn} architectures
%and more complex image classification tasks 
%show less convincing results.
It is envisaged that in the cases where $t$-softmax 
fails to outperform the baseline,
further modifications of the \gls{nn} 
architectures and
better understanding of the effect of the parameter $\nu$
could lead to more positive results. 
%Future research should focus on these directions 
%and on the use of other distributions as well.

\bibliography{library.bib}

\end{document}